\documentclass[12pt]{article}
\usepackage{amsmath,amssymb}
\newcommand{\beq}{\begin{equation}}
\newcommand{\eeq}{\end{equation}}
\newcommand{\ba}{\begin{array}}
\newcommand{\ea}{\end{array}}
\newcommand{\bea}{\begin{eqnarray}}
\newcommand{\eea}{\end{eqnarray}}
\newcommand{\bean}{\begin{eqnarray*}}
\newcommand{\eean}{\end{eqnarray*}}

\newtheorem{theorem}{Theorem}[section]
\newtheorem{prop}[theorem]{Proposition}
\newtheorem{lemma}{Lemma}

\newenvironment{pf}{\begin{proof} \rm}{\hfill$\square$ \end{proof}}
\newcounter{appendix}
\newcommand{\newappendix}[1]{\vspace{10mm}\pagebreak[3]
\addtocounter{appendix}{1}
\renewcommand{\theequation}{\Alph{appendix}.\arabic{equation}}
\setcounter{equation}{0}
\begin{flushleft}{\Large\bf Appendix \Alph{appendix} #1}
\end{flushleft}\nopagebreak\medskip\nopagebreak}
\newcommand{\CH}{{\cal H}}

\newcommand{\CN}{{\cal N}}

\makeatletter
\@addtoreset{equation}{section}
\renewcommand{\theequation}{\thesection.\arabic{equation}}
\makeatother

\def\be{\beta}
\def\al{\alpha}

\def\la{\lambda}

\newcommand{\cmp}[3]{Comm. Math. Phys. {\bf #1} (#2), #3}

\newcommand{\lmp}[3]{Lett. Math. Phys. {\bf #1} (#2), #3}

\newcommand{\jmp}[3]{J. Math. Phys. {\bf #1}, #2, (#3)}
\newcommand{\rref}[1]{(\ref{#1})} 

\newcommand{\wid}[1]{{\widetilde{#1}}}

\def\back{\!\!\!\!\!\!}
\def\gsig{{\mathfrak{g}_\sigma}}

\def\dpt#1#2{{\frac{\partial #1}{\partial t_{#2}}}}
\def\ddd#1#2{\displaystyle{\frac{\partial #1}{\partial #2}}}

\newcommand{\Ha}[1]{H^{(#1)}}

\def\Nij{Nijenhuis}

\def\endpf{\begin{flushright}$\square$\end{flushright}}

\def\alg{{\mathfrak g}}

\def\bih{biham\-il\-tonian}

\def\dncoo{Darboux-Nijenhuis coordinates}

\begin{document}
\begin{flushright}
Ref. SISSA 45/2003/FM
\end{flushright}
\begin{center}
{\Large\bf Gaudin Models and Bending Flows:}\\
{\Large\bf a Geometrical Point of View}
\end{center}
\vspace{0.8truecm}
\begin{center}
{\large
Gregorio Falqui and Fabio Musso\\
SISSA, Via Beirut 2/4, I-34014 Trieste, Italy\\
falqui@sissa.it, musso@sissa.it}\\
\today
\end{center}
\vspace{0.1truecm}
\begin{abstract}\noindent
In this paper we discuss the \bih\ formulation of the (rational XXX) 
Gaudin models of
spin--spin interaction, generalized to the case of $sl(r)$--valued ``spins''.
In particular, we focus on the  
homogeneous models. 
We find a pencil of Poisson brackets that recursively define a complete set of
integrals of the motion, alternative to the set of integrals associated with 
the ``standard'' Lax representation of the Gaudin model. These integrals, 
in the case of $su(2)$, coincide wih the Hamiltonians of the ``bending flows'' 
in the moduli space of polygons in Euclidean space introduced 
by Kapovich and Millson.
We finally address the problem of separability of these flows and 
explicitly find separation coordinates and separation relations for 
the $r=2$ case.
\end{abstract}
\begin{flushleft}
AMS Classification numbers: 70H06, 37K10,
70H20.
\end{flushleft}
\section{Introduction}
In  \cite{Gaudin1}, M. Gaudin proved the integrability of $N$-site  
$su(2)$ (quantum) spin Hamiltonians of the form
\begin{equation}
{\cal{H}}=\sum_{j<l=1}^N 
\frac{c_j-c_l}{a_j-a_l} \vec{\sigma}_j \cdot  
\vec{\sigma}_l, \label{gau}
\end{equation}
where $\vec{\sigma}=[\sigma_x,\sigma_y,\sigma_z]$ are Pauli matrices.
This property follows from the fact that 
one can write the Gaudin Hamiltonian ${\cal{H}}$ as
\begin{equation}
{\cal{H}}=\sum_{j=1}^N c_j H_j\qquad 
\text{ with }  \qquad H_j=\sum_{l \neq j}
 \frac{ {\vec{\sigma}}_j \cdot  {\vec{\sigma}}_l}{a_j-a_l}
\label{integrals}
\end{equation}
and check that the $H_j$'s  define a set of $N-1$ commuting observables.
Since fixing the values of the $N$ Casimirs
$C_j=| \vec{\sigma}_i|^2 $  the system has $N$ degrees of freedom,
the $N-1$ quantities $H_j$, together with, e.g.,
$S_z=\sum_{i=1}^N \sigma_i^z$, provide a complete set of mutually
commuting observables. 

A relevant member of this class of Hamiltonians is obtained 
when one chooses
$c_k$ to be proportional to $a_k$ for all $k$'s, so that, up to a
rescaling, the  Hamiltonian~\rref{gau} becomes 
\begin{equation}
{\cal{H}}=\sum_{j,l=1}^N {\vec{\sigma}}_j \cdot  {\vec{\sigma}}_l.
\label{gaudeg}
\end{equation} 
This is the Hamiltonian of the XXX rational homogeneous Gaudin model.

This system is not only integrable, 
but maximally superintegrable. One can understand this stronger property
as follows (see, e.g \cite{HaYe03}): 
since the ``physical'' 
Hamiltonian (\ref{gaudeg}) is independent of the parameters, the choice 
of the $a_k$'s in the definition of the
commuting integrals $H_j$ is arbitrary (provided
$a_i \neq a_j,\>i \neq j$). 
So, choosing another set of parameters
$b_k\neq a_k$ and considering
$\tilde{H}_l=\sum_{l \neq j} \frac{ {\vec{\sigma}}_j \cdot
  {\vec{\sigma}}_l}{b_j-b_l}$
one can  define
the two sets of complete commuting quantities:
\begin{displaymath}
\{ {\cal{H}}, H_1, \dots, H_{N-2}, S_z \}; \quad   
\{ {\cal{H}}, \tilde{H}_1, \dots, \tilde{H}_{N-2}, S_x \}.
\end{displaymath}
Since for generic choices of the sets $a_k,b_k$ the observables
\[
\{\CH, H_1, \dots, H_{N-2},\tilde{H}_1, \dots, \tilde{H}_{N-2},S_z,S_x\}
\] 
are algebraically independent,  the model is indeed
maximally superintegrable.

Recently it was pointed out independently by various authors \cite{Ragn,Kar}, 
that with (\ref{gaudeg}) it is possible to associate 
a set  of commuting integrals {\em independent} of the parameters.
Such operators are of the form:
\begin{equation}
I_{k-1}=\sum_{j,l=1}^{k} {\vec{\sigma}}_j \cdot  {\vec{\sigma}}_l  
\quad k=2,\dots, N, \label{whatever}
\end{equation}
and, together with,  $S_z$ they form
a complete set of involutive integrals for $\CH$.

In a completely different context the classical version 
of the integrals (\ref{whatever}), namely
\begin{equation}\label{pind}
K_j=\frac{1}{2} \text{Tr}\Big( \big( \sum_{i=1}^{j+1} A_i \big)^2 \Big), 
\end{equation}
where $A_1, \ldots A_N$ are generic elements of $su(2)$,
were considered by Kapovich and Millson~\cite{KM}. 
These authors (see also
\cite{FlMi01}) studied the moduli space of $N+3$-sided polygons in
$\mathbb{R}^3$, and (implicitly) showed  
that it coincides with a suitable Marsden-Weinstein
quotient (with respect to the diagonal action of 
$SU(2)$) of the phase 
space of the $N+3$ site $su(2)$--Gaudin models. They remarked that such a
space possesses a natural Hamiltonian structure, and integrated, via
action-angle variables methods, the flows associated with  
the integrals $K_l,l=1,\ldots,N$.  
It is worthwhile to remind the intriguing representation
of such flows: if one draws, from a chosen vertex, the $N$ possible
diagonals of an $N+3$-sided polygon, the flow associated with the Hamiltonian  
$K_k$ geometrically represents the bending of one side of the polygon along
the $k$-th diagonal (the other side being kept fixed), 
whence the name of ``bending flows''. 

The Gaudin system~\rref{gau} admits various generalizations.
Gaudin himself pointed out that the integrals (\ref{integrals}) can be
generalized to any semisimple Lie algebra $\alg$. Clearly, if 
the rank of $\alg$ is greater than $1$, the number of such integrals is not
enough to ensure complete integrability. 
The missing integrals have been shown by Jur\v{c}o \cite{Jurco} and
Sklyanin~\cite{Sk92}
to be provided by the spectral invariants of a suitable  Lax matrix, whose
classical counterpart is
\begin{equation}
L_{rat}=\sum_{i=1}^N \frac{A_i}{\lambda-a_i}, \label{Lrat} 
\end{equation}
where
$a_i\neq a_j, \> 
i\neq j$ and the $A_i$ are generic elements of $\alg$.
In terms of the Lax matrix (\ref{Lrat}) the generalization of 
the Hamiltonian (\ref{gaudeg}) reads:
\begin{equation}
H_G=\sum_{i=1}^N res_{\vert_{\lambda=a_i}} \text{Tr} 
(\lambda L_{rat}^2)= \sum_{j \neq i} \text{Tr}(A_i A_j). \label{HG}
\end{equation}

Another straightforward generalization of this model
is obtained adding a 
constant term $\sigma$ to the Lax matrix.
In the $su(2)$ case,
this is equivalent to adding to 
the Hamiltonians (\ref{gaudeg}) a term describing
the interaction of the spins with a magnetic field with a constant direction
in each site but with different intensity. In this case one speaks of
``inhomogeneous'' Gaudin magnet. The complete integrability
and separability of these systems, (for the $\alg=sl(n)$ case) was studied and
proved in \cite{Sk89,Gekhtman,Scott}.

The aim of this paper is to frame the analysis of the  Gaudin
models, as well as of the Hamiltonians~\rref{pind} of the
bending flows of Kapovich and Millson,
in the scheme of \bih\ geometry as advocated by Gel'fand and Zakharevich
\cite{GZ}, and to show how one can
use this scheme to explicitly integrate the model for $\alg=sl(2)$.
We will consider only the classical models, and consider the complexified
case (that is, we will study the Gaudin system associated with a complex
semisimple Lie algebra $\alg$).

Our first task will be to briefly show how, using nowadays standard results
of the theory of $r$-matrices on loop algebras~(see, e.g., \cite{RSTS}), 
one can provide the phase space of 
the (inhomogeneous) Gaudin magnet with a \bih\ structure, selecting it
out of a   multi-parameter family of Poisson structures.
This  structure gives rise, 
according to the GZ scheme, to the integrals associated with the Lax 
matrix of Jur\v co and Sklyanin~\rref{Lrat}. 

Then we will construct, in the homogeneous case, 
{\em another} \bih\ structure, non compatible (in a sense to be precised
later) with the abovementioned family, whose GZ analysis gives rise, 
in the $sl(2)$ case, to the parameter independent 
integrals \rref{pind}. Since such additional \bih\ structure is still
constructed within a Lie--theoretical setting, we will be able 
to straightforwardly apply this scheme to $\alg=sl(r)$, with arbitrary $r$.
In this way, we will be able   
to find  a sufficient  number of commuting integrals to be added to 
the  ``generalized bending Hamiltonians'' $I_k$, 
yielding a complete family of integrals  alternative 
to the ``standard'' family obtained by Sklyanin and Jur\v{c}o.

The GZ analysis of such a model will finally lead us to introduce a kind of
Lax matrices for such flows and to show that the Hamilton--Jacobi equations
associated with the $sl(2)$ bending Hamiltonians are separable by computing 
explicitly the separation variables and the separation relations.

\section{GZ analysis of Gaudin models}\label{sec:general}

The Gel'fand--Zakharevich (GZ) scheme \cite{GZ} for integrating a 
\bih\ system can be seen as a particularly efficient scheme to implement the
Lenard--Magri recursion for manifolds endowed with a pair of compatible Poisson
brackets none of which is symplectic (i.e., non--degenerate).

One considers a manifold $M$ endowed with a pair of compatible
Poisson tensors $P_1-\la P_0$, or, in other words, a pencil of 
Poisson brackets
\begin{equation}
  \label{eq:g1}
  \{f,g\}_\la=\{f,g\}_{P_1}-\la\{f,g\}_{P_0}=
\langle df, (P_1-\la P_0) dg\rangle.
\end{equation}
(where $\langle\cdot,\cdot\rangle$ 
is the canonical pairing between $T^*M$ and $TM$), and assumes that
the kernel of the generic element of the Poisson pencil be $k$--dimensional. 
Let $C_1,\ldots,C_k$ be independent Casimir 
functions of $P_0$. The GZ method,
roughly speaking, suggests to use these Casimirs as ``starting'' elements for
Lenard chains yielding (under some technical additional conditions), 
via the  method of \bih\ iteration, families of functions
$\{\Ha{a}_m\}_{a=1,\ldots,k}^m=0,\ldots$, such that
for any function $F$ on $M$, and $a=1,\ldots, k,$
\begin{equation}\label{eq:lenrecrel}
\{F, \Ha{a}_m\}_{P_0}=\{F,\Ha{a}_{m-1}\}_{P_1}, 
\quad \text{ with } \Ha{a}_0=C_a.
\end{equation}
 As a consequence of the \bih\
iterative scheme and of the fact that all Lenard chains start with a Casimir
function of $P_0$ (they are ``anchored'', in the language of~\cite{GZ}), all
these functions are mutually in involution with respect to 
both Poisson brackets.
Obviously, the maximal number of independent functions one may hope to
get in this way is $N_{max}=\frac12(\text{dim}M+k)$.  
If this is indeed the case, the 
geometric scheme herewith outlined defines families of completely
integrable systems in the Liouville sense. Indeed, let us suppose that the GZ
method provides us 
with $k$ families of mutually commuting independent 
functions
\[
\{\Ha{a}_m\}_{a=1,\ldots,k}^{m=0,\ldots,n_a},\quad \text{ with } 
\sum_{a=1}^k n_a=\frac12(\text{dim}M+k).
\]
Let $\CH$ be a generic
 element in the ring generated by such commuting functions, 
and let $X_\CH=P_0 d\CH$ be the corresponding Hamiltonian vector field. 
Let us consider a generic symplectic leaf $S\subset M$ of $P_0$; it is a
$d_s=\text{dim}M-k$ dimensional manifold, with the natural symplectic form
induced by the Poisson structure $P_0$. $X_\CH$ clearly 
restricts to $S$, and, as a 
consequence of the \bih\ iteration on $M$, comes equipped with 
$\frac12(\text{dim}M+k)-k=\frac12 d_S$ integrals in involution, given by the
restriction to $S$ of the functions 
$\{\Ha{a}_m\}_{a=1,\ldots,k}^{m=1,\ldots,n_a}$. As a consequence of the
genericity assumption on the symplectic leaves, these functions will be
independent on $S$ as well and give the complete family of involutive
integrals required by the Liouville theorem.

The aim of this Section is 
to frame the (general, that is, inhomogeneous) Gaudin model
within the \bih\ scheme, and to reinterpret its complete integrability within
the theoretical framework of the GZ analysis briefly sketched above. The
manifolds we will consider will be Cartesian products of a Lie algebra 
$\alg^N$, and the Poisson pencils suitable {\em linear} pencils of
$\alg^N$. Since these results are essentially known in the literature, 
we will mostly limit ourselves to state results and sketch proofs, 
referring to~\cite{RSTS,PV} for a more general setting, and 
to \cite{FM1} for the explicit study of the 3-particle $sl(2)$-case.

\subsection{Notations and Conventions}
Let us briefly recall the notion of Lie-Poisson brackets associated with a Lie 
algebra and fix some notations and conventions we will use throughout 
the paper. 

If $\alg$ is a Lie algebra, its dual
$\alg^*$ comes equipped with 
the standard Lie-Poisson structure:
\begin{equation}
\{ F, G \}(A)= \langle A, [d F, d G]\rangle =\langle dF,P dG \rangle,
\quad F,G\> \in C^\infty(\alg^*). 
\label{LPB}
\end{equation}
If $\alg$  is semisimple
we can identify $\alg^*$ with $\alg$. Indeed, 
we can associate a matrix $X_A$ with
any element $A\in \alg^*$.  considering, e.g.,
the fundamental  representation of the algebra $\alg$, 
and taking the trace form as a  
bilinear non-degenerate pairing
\[ 
\langle A, X_B\rangle=\text{Tr}(X_A\cdot X_B).
\]
From now on we will implicitly use this identification, and write $A,B\ldots$ 
instead of $X_A, X_B,\ldots$ for simplicity of notation.
Using the ciclicity of the trace, the Hamiltonian vector field associated 
by (\ref{LPB}) with a smooth function $F$ is represented by:
\begin{displaymath}
PdF=\dot{A}=\left[ A , \frac{\partial F}{\partial A} \right],
\end{displaymath}  
where the symbol $\frac{\partial
  F}{\partial A}$ denotes the matrix satisfying, for any  $\Xi$
in $\alg$, 
\[
F(A+t\Xi)=F(A)+t\cdot\text{Tr}\,(\ddd{F}{A}\cdot\Xi)+o(t).
\]
If we take the direct product of $N$ copies 
of $\alg$, the standard Lie-Poisson structure
becomes:
\begin{equation}
\{ F, G \}(A_1, \dots,A_N)= \sum_{i=1}^N \langle A_i, 
\left[ \frac{\partial F}{\partial A_i}, 
\frac{\partial G}{\partial A_i}\right] \rangle \label{LPB2}
\end{equation}   
and the Hamiltonian vector field associated with 
a function $F$ is
\begin{displaymath}
\dot{A_i}=\left[ A_i , \frac{\partial F}{\partial A_i} \right],\quad
i=1,\ldots, N.
\end{displaymath} 
We can write the above equation in the form:
\begin{equation}\label{eq:sumrep}
\dpt{A_i}{}=(X_F)_i= (P d F)_i = \sum_{j,k} p_{ijk} 
\left[ A_k,\frac{\partial F}{\partial A_j} \right]  \text{ with }
p_{ijk}= \delta_{ij} \delta_{jk}.
\end{equation}
We will also often write $P$ (and other Poisson tensors)  
representing its action on the
differential of a function by means of the matrix symbolic form: 
\begin{equation}
\left(\begin{array}{c}\dot{A}_1\\ \dot{A}_2\\ \vdots \\
\dot{A}_N\end{array}\right)=
\left( 
\begin{array}{cccc}
[A_1,.] & 0 & \dots & 0\\
0 & [A_2,.] & \dots & 0\\
\vdots & \vdots &  & \vdots\\
0 & 0 & \dots & [A_N,.]  
\end{array}
\right) \cdot
\left(\begin{array}{c}\ddd{F}{A_1}\\\ddd{F}{{A}_2}\\ \vdots \\
\ddd{F}{{A}_N}\end{array}\right).
\label{PDA}
\end{equation}
For this reason, we will term the standard Lie-Poisson tensor
$P$ on $\alg^N$ the {\em diagonal} Poisson tensor.

\subsection{A Bihamiltonian Structure of the Gaudin model}
A bihamiltonian structure for rational Gaudin models can be obtained 
using the following argument.
Let us consider the map 
$\{ A_i \} \longrightarrow \{ B_i \}$
that sends the rational Lax matrix 
\begin{displaymath}
L_{rat}=\sigma+\sum_{i=1}^N \frac{A_i}{\la-a_i} 
\end{displaymath}
in the polynomial Lax matrix 
\begin{equation}
L_{poly}=\la^n\sigma+\sum_{i=0}^{N-1} B_i \lambda^i
=\big(\prod_{i=1}^N (\lambda-a_i)\big)\cdot L_{rat}
 \label{Lpoly} 
\end{equation}
given explicitly by
\begin{equation}
\label{map}
\begin{split}
B_l&=(-1)^{N-l-1} \sum_{i=1}^N  s_{N-l-1}
(a_1, \dots, \hat{a_i}, \dots, a_N)\cdot A_i+  \\
&(-1)^{N-l}s_{N-l}(a_1, \dots, a_N)\cdot  \sigma,\qquad 
l=0,\dots,N-1, 
\end{split}
\end{equation}
where $s_k(a_1,\dots, a_N)$ denotes the $k$-th elementary 
symmetric polynomial in the variables $a_1,\dots, a_N$.

On the space of polynomial pencils of matrices a family of
mutually compatible Poisson brackets are defined~\cite{RSTS,MaMo84}. 
They will be termed, for the sake of brevity, RSTS tensors. 
In a nutshell, this family can be described  by saying that there is a 
map from degree $N$ polynomials  in the variable $\la$
to the set of Poisson structures on the manifold of polynomial Lax matrices
of the form~\rref{Lpoly} which sends the monomials $\la^0, \ldots, \la^N$
into $N+1$ fundamental Poisson brackets, $\Pi_l, l=0, \ldots, N$. 
In our case, the fundamental tensors $\Pi_l$ can be represented by 
matrices having the following 
block--diagonal structure:
\begin{equation}
\Pi_l= 
\left( 
\begin{array}{cc}
C_l & 0 \\
0 & D_l 
\end{array}
\right), \label{PRSTS}
\end{equation}
with:
\begin{eqnarray}
&& \left\{           
\begin{array}{l}
(C_l)_{ij}= -[B_{i+j-l-1},.] \qquad i,j=1, \dots, l  \\
(D_l)_{ij}= [B_{i+j+l-1},.]  \qquad i,j=1, \dots, N-l
\end{array}
\right. \\
&& B_i=0 \  {\rm{if}}  \ i <0  \ {\rm{or}}  \ i>N, \qquad\text{ and}  
\quad B_N=\sigma. 
\nonumber 
\end{eqnarray}

\begin{lemma} In the ``coordinates'' $B_0,\dots,B_{N-1},\sigma$, the diagonal Poisson tensor 
$P$~\rref{PDA} is 
given by the sum
\begin{equation}
P=\sum_{l=0}^N (-1)^{N-l-1} s_{N-l}(a_1, \dots,
a_N) \Pi_l,  \label{PDB}
\end{equation}
where the $s_i$'s are the elementary symmetric polynomials in the $a_i$'s,
that is, it is the tensor associated with the polynomial
\[
p_N=\prod_{l=1}^N (\la-a_i).
\]
\end{lemma}
This Lemma can be proved by means of a direct computation.
For the reader's convenience, we collect the main steps of it in Appendix A.

Since the Poisson tensors (\ref{PRSTS}) form a $N+1$-parameter family of
compatible Poisson tensors, we can choose as a second Poisson tensor 
a suitable linear combination of them to have a \bih\ structure on $\alg^N$.
Let
\begin{equation}
Q=\sum_{l=0}^{N-1} (-1)^{N-l} s_{N-l-1}(a_1, \dots,
a_N) \Pi_l,  \label{P1}
\end{equation}
be the tensor associated with the polynomial
\[
p_{N-1}=\big(\frac{p_N}{\la}\big)_+
=\la^{N-1}-s_1\la^{N-2}+\cdots+(-1)^{N}s_{N-1}.
\] 
All the integrals of motion that one can obtain from
the spectral invariants of the Lax matrix \rref{Lrat} can be
obtained by the GZ method applied to the pencil $Q-\la P$; 
in fact it holds (see, also, \cite{RSTS}):
\begin{lemma}
All the vector fields associated to the spectral invariants of 
\rref{Lrat} are bihamiltonian with respect to the pair $Q-\la P$.
\end{lemma}
\noindent{\bf Proof:} We find convenient to 
work in the variables $B_i$. 
Let us define:
\begin{eqnarray}
&& K^{(i)}_\al=\text{Tr}\big(\text{Res}_{\lambda=0} 
\big( \frac{(\sum_{j=1}^N B_j \lambda^j)^\al}{\lambda^i} \big) \big) 
\label{spec} \\
&& i=1, \dots, \al N \qquad \al=2,\dots \text{rk}(\alg).\nonumber
\end{eqnarray} 
For any fixed $\al$, the $\al N$ functions \rref{spec} fulfill the 
relations \cite{RSTS}:
\begin{equation}
\Pi_i d K^{(j)}=\Pi_{i+k} d K^{(j+k)}=X^{(j-i)}. \label{rec2}
\end{equation}
{}From \rref{rec2} it follows that $X^{(i)}=0$ 
if $i \leq 0$ or $i>N(\al-1)$; in fact, if
$i \leq 0$ then $K^{(i)}=0$ and $X^{(i)}=\Pi_0 d K^{(i)}=0$, while if $i>N(\al-1)$,
then $K^{(N+i)}=const$ and $X^{(i)}=\Pi_N d K^{(N+i)}=0$. 
Now let us set:
\begin{equation}
b_l=(-1)^{N-l+1} s_{N-l}(a_1,\dots,a_N), \label{short}
\end{equation}
we have:
\begin{eqnarray*}
&& P d K^{(j)}=\sum_{l=0}^{N} b_{l} \Pi_l d K^{(j)}= \sum_{l=0}^{N} b_{l} X^{(j-l)}\\
&& Q d K^{(j)}=\sum_{l=1}^{N} b_{l} \Pi_{l-1} d K^{(j)}= 
\sum_{l=1}^{N} b_{l} X^{(j-l-1)}.
\end{eqnarray*}
Then:
\begin{displaymath}
P d K^{(j)}_\al - Q d K^{(j+1)}_\al= b_0 X^{(j)}_\al.
\end{displaymath}
If one of the $a_i$ is equal to zero, 
then $b_0=\prod_{i=1}^N a_i=0$ and the proof is concluded. 
Otherwise we need to find a function $F^{(j)}_\al$ such that
\begin{displaymath}
Q d F^{(j)}_\al=b_0 X^{(j)}_\al
\end{displaymath}
We proceed by induction. If $j=1$, we have $
b_0 X^{(1)}_\al =Q \frac{b_0}{b_1}d K^{(1)}_\al$, so that
$F^{(1)}_\al =b_0/b_1 K^{(1)}_\al$. 
Now let 
$F^{(i)}_\al$ 
be such that:
$b_0 X^{(i)}_\al=Q dF^{(i)}_\al \qquad i=1,\dots, j-1$.
Then
\begin{eqnarray*}
&& Q \frac{b_0}{b_1} d K^{(j)}_\al=b_0 X^{(j)}_\al +\frac{b_0 b_2}{b_1}X^{(j-1)}_\al+ \dots 
+\frac{b_0 b_N}{b_1}X^{(j-N+1)}_\al \Longrightarrow \\
&& \Longrightarrow b_0 X^{(j)}_\al=Q \left( \frac{b_0}{b_1} d K^{(j)}_\al- 
\frac{b_2}{b_1} d F^{(j-1)}_\al-\dots- \frac{b_N}{b_1} d F^{(j-N+1)}_\al \right).
\end{eqnarray*}
So we have:
\begin{equation}
Q dF^{(j)}_\al=b_0 X^{(j)}_\al,\>\text{ with }\> F^{(j)}_\al= \frac{b_0}{b_1} K^{(j)}_\al- 
\frac{1}{b_1}\sum_{i=1}^{N-1} b_{i+1} F^{(j-i)}_\al
\end{equation}
\endpf
Some observations on the GZ sequences are in order. The starting
points of the GZ sequences are given by the Casimirs of $P$. 
We have to distinguish two cases:
\begin{description} 
\item[a)] If $b_0 \neq 0$, i.e. all 
the $a_i$ are different from zero, then the Casimirs 
of $P$ are given in terms of the spectral invariants \rref{spec} by the following 
expressions:
\begin{equation}
C_{i,\al}=\sum_{j=1}^{\al N} a_i^{j} K^{(j)}_\al \qquad i=1,\dots,N \qquad \al=2,\dots
\text{rk}(\alg). \label{CKi} 
\end{equation}  
For any $\al$, starting the GZ sequences from suitable linear combinations of the 
Casimirs $C_{i,\al}$ we can construct $N$ GZ sequences of length $\al-1$ 
(i.e. defining $\al-1$ independent vector fields) each 
starting with a Casimir of $P$ and ending with a Casimir of $Q$.
\item[b)] If $b_0=0$ then only one among the $a_i$, say $a_N$, is zero. 
In this case equation \rref{CKi} defines
$(\text{rk}(\alg)-1)(N-1)$ independent Casimirs, instead of $(\text{rk}(\alg)-1) N$:
\begin{equation}
C_{i,\al}=\sum_{j=1}^{\al N} a_i^{j} K^{(j)}_\al \qquad i=1,\dots,N-1 \qquad \al=2,\dots
\text{rk}(\alg). \label{CKi2} 
\end{equation}     
The functions \rref{CKi2} turns out to be simultaneous 
Casimirs for both $P$ and $Q$.
The remaining $\text{rk}(\alg)-1$ Casimirs 
of $P$ (the rank of $P$ is obviously the 
same in both cases) are given by 
\begin{equation}
C_{N_\al}=K^{(1)}_\al \qquad \al=2,\dots,\text{rk}(\alg). \label{CN}
\end{equation}
To each Casimir \rref{CN} is associated a GZ sequence of length $(\al-1)N$.
\end{description}
\section{The homogeneous case}\label{sec:homo}
The  constant term $\sigma$ in the Lax matrix~\rref{Lrat}
physically describes the coupling of the $i$--th spin
with an  ``external magnetic'' field $\beta_i=a_i\, \sigma$.
The matrix $\sigma$ in the definition of 
the rational Lax matrix~\rref{Lrat}
is somewhat a free parameter in the theory. Changing $\sigma$ amounts to ``changing the
direction'' of this magnetic field.
The choice usually done in the literature is the generic 
one (say, $\sigma$ is a diagonal matrix with
different entries); this ensures the functional independence 
of the coefficients
of the spectral invariants of $L_{rat}$, whence the fact that
they are in a sufficient number to yield complete integrability of the model.

If $\sigma$ is not generic,
but the dimension of its stabilizer
 $ \mathfrak{g}_\sigma:=\{\tau\in \mathfrak{g} \text{ s.t. } [\tau,\sigma]=0\}$
is greater than the rank of $\alg$ the following happens. Not all the spectral
invariants of the Lax matrix are functionally independent, but one can recover
the ``missing'' integrals noticing that the functions:
\begin{displaymath}
F_\tau=\text{Tr} \Big( \sum_{i=1}^N A_i \tau \Big), \quad \tau\in\gsig,
\end{displaymath}
commute with all the spectral invariants of $L_{rat}$. 

However, something more substantial
occurs for $\sigma=0$, that is, in the homogeneous case.
As we have recalled in the Introduction,
in such a case $H_G$ (\ref{HG}) defines, for $\alg=sl(2)$,
a superintegrable Hamiltonian system and, in particular, 
it is possible to find another complete set of commuting 
first integrals which are not explicitly dependent on  
the parameters $a_i$.

From now on we will focus on this additional family of integrals, that, in
the classical $N$-site $sl(2)$ model 
can be given by the very simple formula:
\begin{equation}
  \label{eq:rintcl}
I_{l-1}=\sum_{j,k=1}^l \text{Tr}(A_j\cdot A_k), l=2,
\ldots, N.  
\end{equation}
We will introduce a
further Poisson structure $R$ on the manifold $\alg^N$. As we shall show 
it will be possible to combine it with the diagonal 
Poisson structure $P$ of Eq.~\rref{PDA} to get
a further {\em Poisson pencil}, not belonging to the RSTS family described 
in Section \ref{sec:general}. The  GZ method applied to the Poisson pencil 
$R-\la P$ will give rise to these new set of integrals. 
Since everything will be done in a
Lie-algebraic setting, these results hold for a generic semisimple Lie
algebra, and in particular, for $sl(r)$ with arbitrary $r$.
\subsection{The additional bihamiltonian pencil}
Let us consider the bivector $R$, defined, by means of the constructions
outlined in Section~\ref{sec:general} by the following matrix:
\begin{equation}
R=\left(
\begin{array}{cccc}
0 & \left[ A_1,\cdot \right]  & \cdots & [A_1,\cdot]\\
\left[ A_1, \cdot \right] & [A_2-A_1,\cdot]  & \cdots & [A_2,\cdot]\\
\vdots  & \vdots & \ddots & \vdots \\
\left[ A_1,\cdot \right] & [A_2,\cdot]  & \cdots & [(N-1)A_N-
\sum_{i=1}^{N-1} A_i,\cdot]
\end{array} 
\right). \label {P2}
\end{equation}

\begin{prop}\label{prop:rpoiss}
The bivector $R$ defined by \rref{P2} is a Poisson bivector, and it is
compatible with the diagonal Poisson tensor $P$.
\end{prop}
{\bf Proof:} Linearity and antisymmetry are obvious, so we must prove only the 
Jacobi identity. Also, we can limit ourselves to prove the assertions for 
the case of linear functions on $\alg$. 
If $F,G,H$ are such functions, identifying their
differentials with the three $N$--tuples of matrices
$\{\alpha_i\},\{\beta_i\},\{\gamma_i\}$, (e.g.,
$\ddd{F}{A_i}=\alpha_i,\ldots$ ), 
the Poisson bracket is defined by:
\begin{displaymath}\begin{split}
\{ F, G \}_R&= \langle dF , R dG \rangle= \sum_{i,j,k} r_{ijk} \text{Tr} 
\left(\alpha_i\left[ A_k,
\beta_j\right] \right)=\sum_{i,j,k} r_{ijk} \text{Tr}
\left(A_k \left[ \beta_j,
\alpha_i\right] \right), \\ &
r_{ijk}=(k-1) \delta_{ij} \delta_{jk}-\theta_{(i-k)} \delta_{ij} +\theta_{(j-i)}
\delta_{ik}+\theta_{(i-j)}\delta_{jk},
\end{split}
\end{displaymath}
where $\delta$ is the usual Kronecker symbol and $\theta$ is the discrete
Heaviside function, normalized as
\begin{displaymath}
\theta_{(i)}= \left\{\begin{aligned}
& 1 \qquad \text{if} \ i >0 \\
&0 \qquad \text{if} \ i \leq 0
\end{aligned}
\right. 
\end{displaymath}
The Jacobi identity reads 
\begin{equation*}\begin{split}
 &\qquad\qquad\{ H, \{ F, G \}_R \}_R +  \{ F, \{G,H \}_R\}_R+ \{ G, \{ H,F \}_R \}_R
 =\\
&  \sum_{i,j,k,l,m} r_{ijk} r_{lmj}\left( \text{Tr}\left( A_k \left[ \left[\beta_m,\alpha_l\right],\gamma{_i} \right]
\right)+ \text{Tr}\left( A_k \left[ \left[
\alpha_{m},
\gamma_{l}\right],\beta_{i} \right]
\right) + \text{Tr}\left( A_k \left[ \left[
\gamma_{_m},
\beta_{l}\right],\alpha_{i} \right]
\right) \right),\end{split}
\end{equation*}
which, renaming the indices, becomes
\begin{equation*} \begin{split}
&\sum_{i,j,k,l,m} r_{ijk} r_{lmj} \text{Tr}\left( A_k 
\left[ \left[ 
\beta_{_m},\alpha_{l}\right],\gamma_i \right]
\right)+r_{mjk} r_{ilj} \text{Tr}\left( A_k \left[ \left[
\alpha_{l},
\gamma_{i}\right],\beta_{m} \right]
\right)+\\ 
&\qquad +r_{ljk} r_{mij} \text{Tr}\left( A_k \left[ \left[
\gamma_{i},
\beta_{m}\right],\alpha_{l} \right]
\right). \end{split}
\end{equation*} 
A sufficient condition for the last expression to be zero is that for any $k$ it holds:
\begin{equation}
\sum_j r_{ijk} r_{lmj}= \sum_j r_{mjk} r_{ilj} \label{cyclic}
\end{equation}
In fact, a consequence of \rref{cyclic} to hold is that
implies that $ t_{iklm}=\sum_j r_{ijk} r_{lmj}$ 
is invariant for cyclic permutations of the indices $i,l,m$.
So, if
\rref{cyclic} holds we can write:
\begin{eqnarray*}
&& \back \back \{ H, \{ F, G \}_R \}_R +  \{ F, \{G,H \}_R\}_R+ \{ G, \{ H,F \}_R \}_R
 =\\
&& \back \back  = \sum_{i,k,l,m} t_{iklm} \text{Tr}\left( A_k 
\left( \left[ \left[ 
\beta_{_m},\alpha_{l}\right],\gamma_i \right]+\left[ \left[
\alpha_{l},\gamma_{i}\right],\beta_{m} \right]+ \left[ \left[
\gamma_{i},
\beta_{m}\right],\alpha_{l} \right]
\right) \right),
\end{eqnarray*}
which vanishes thanks to the Jacobi identity in $\alg$.

Let us show that  \rref{cyclic} holds in our case.
By means of algebraic manipulations, namely cycling through 
$i,l,m$ and renaming the indices using the Kronecker's 
$\delta$, we obtain:
\begin{eqnarray*}
&& \back \back \sum_j r_{ijk} r_{lmj}- \sum_j r_{mjk} r_{ilj}=\\
&& \back \back =\delta_{ik} \delta_{lm} [(l-i) \theta_{(l-i)} \sum_j \theta_{(j-i)} \theta_{(l-j)}]+ 
 \delta_{il} \delta_{mk} [(i-m) \theta_{(i-m)}+ 
\sum_j \theta_{(j-m)} \theta_{(i-j)}]+\\
&& \back \back + \delta_{lm}[\theta_{(i-k)} (\theta_{(l-i)}- \theta_{(l-k)})+
\theta_{(l-k)} \theta_{(i-l)}]+ 
\delta_{il} [\theta_{(m-k)}( \theta_{(i-k)}-\theta_{(i-m)})- \theta_{(i-k)}
\theta_{(m-i)}]+\\
&& \back \back + \delta_{ik}[\theta_{(l-i)} (\theta_{(m-l)}- \theta_{(m-i)})+
\theta_{(m-i)} \theta_{(l-m)}]+
\delta_{mk}[\theta_{(l-m)} (\theta_{(i-m)}- \theta_{(i-l)})-
\theta_{(i-m)} \theta_{(l-i)}]
\end{eqnarray*}
Using the identities:
\begin{equation*}\begin{split}
&\sum_j \theta_{(j-i)}\theta_{(l-j)}=(l-i-1) \theta_{(l-i)}\\
& \theta_{(i-k)} (\theta_{(l-i)}- \theta_{(l-k)})+
\theta_{(l-k)} \theta_{(i-l)}=-\theta_{(l-k)} \delta_{il}
\end{split}
\end{equation*}
we see that every term cancels out.\\
We now prove that $R$ is compatible with the diagonal tensor $P$.
The characteristic  condition for the compatibility
of two Poisson tensors is 
\begin{displaymath}
\{ H, \{ F, G \}_P \}_R+ \{ H, \{ F, G \}_R \}_P +\text {cyclic permutations
  of} F,G,H =0.
\end{displaymath}
Recalling that 
$\{ F, G \}_P=\sum_{i,j,k}\delta_{ij} \delta_{jk}  \text{Tr}
\left(A_k \left[ \frac{\partial G}{\partial A_j},
\frac{\partial F}{\partial A_i}\right] \right)$, one shows that
a sufficient condition for the compatibility of $R$ and $P$
is that the quantity 
\begin{displaymath}
s_{iklm}=\sum_j \left( r_{ijk} \delta_{lm}\delta_{mj}+ 
\delta_{ij}\delta_{jk} r_{lmj} \right)
\end{displaymath}
be invariant under cyclic permutations of the indices $i,l,m$ for all $k$'s.
Actually, 
\begin{equation*}
 s_{iklm}=
(k+i) \delta_{ik} \delta_{lm} \delta_{il}- \theta_{(i-k)}\delta_{lm} 
\delta_{il}+\theta_{(i-m)}\delta_{lm} \delta_{kl}+\theta_{(m-l)}\delta_{ik} 
\delta_{il}+\theta_{(l-m)}\delta_{ik} \delta_{im}
\end{equation*}
that manifestly satisfies this property.
\endpf
{\bf Remark.} By the previous proposition, we can endow, for every $N$, the
space $(\alg^*)^N$ with a \bih\ structure $P_\la=R-\la P$. A natural question
arises, that is what is the connection with the RSTS family of Poisson
structures discussed in Section~\ref{sec:general}. We do not have yet 
a complete answer to this point; however, as we will show in Appendix B,
the new tensor $R$ is {\em not} compatible with the generic element of the
RSTS family~\rref{PRSTS}.

\subsection{The Lenard Chains}\label{subs:lmc}  
We construct the Lenard chains for the Poisson pencil $R-\la P$, using the GZ 
recipe discussed in Section \ref{sec:general}.
To shorten notations we define:
\begin{eqnarray}
&& B_l=\sum_{i=1}^{l-1} A_i \label{Bl}, \quad 
F^{(\alpha)}_{\beta,l}= {\rm res}_{\lambda=0} \frac{1}{\lambda^{\alpha+1}} 
\left( \lambda A_l + B_l \right)^\beta \label{Fm}\\
&& H^{(\alpha)}_{\beta,l}=\text{Tr}(F^{(\alpha)}_{\beta,l}). \label{Hm}
\end{eqnarray} 
We will first find a kind of ``modified'' recursion relation.
\begin{lemma}\label{lemma:4}
It holds:
\begin{equation}
P d H^{(\alpha-1)}_{\beta,l}=(R -(l-2) P) d H^{(\alpha)}_{\beta,l} \label{de}
\end{equation}
for $\alpha=1,\dots,\beta$
\end{lemma}
{\bf Proof:}
We proceed by induction on $\beta$. 
If $\beta=2$ we have:
\begin{eqnarray*}
&&  H^{(0)}_{2,l}=tr(B_l^2) \quad \Rightarrow \quad \frac{\partial H^{(0)}_{2,l}}{ \partial A_j}=2 \theta_{(l-j)}B_l\\
&&  H^{(1)}_{2,l}=2 tr(A_l B_l) \quad \Rightarrow \quad \frac{\partial H^{(1)}_{2,l}}{ \partial A_j}=2 \theta_{(l-j)}A_l+B_l \delta_{jl}\\
&&  H^{(2)}_{2,l}=tr(A_l^2) \quad \Rightarrow \quad \frac{\partial H^{(2)}_{2,l}}{ \partial A_j}=2 A_l \delta_{jl}.
\end{eqnarray*}
By direct computation we obtain:
\begin{eqnarray*}
&& ((R-(l-2)P)dH^{(2)}_{2,l})_i=2 ( \theta_{(l-i)} [A_i,A_l]+ \delta_{il} [A_i,B_i])=(P d H^{(1)}_{2,l})_i\\
&& ((R-(l-2)P)dH^{(1)}_{2,l})_i= 2 \theta_{(l-i)} [A_i,B_l] = (P d H^{(0)}_{2,l})_i
\end{eqnarray*}  
We use the inductive hypothesis in the case $\alpha \leq \beta-1$.
Plugging in the following identities:
\begin{eqnarray}
&& [B_l,F^{(\alpha)}_{\beta-1,l}]+[A_l,F^{(\alpha-1)}_{\beta-1,l}]=0 \label{iid}\\
&& \frac{ \partial H^{(\alpha)}_{\beta,l}}{ \partial A_j}=\theta_{(l-j)} \frac{ \partial H^{(\alpha)}_{\beta,l}}{ \partial B_l}
+ \delta_{jl}  \frac{ \partial H^{(\alpha)}_{\beta,l}}{ \partial A_l}  \label{iid2}\\
&& \frac{ \partial H^{(\alpha)}_{\beta,l}}{ \partial A_j} = \theta_{(l-j)}F^{(\alpha)}_{\beta-1,l}+ B_l 
\frac{ \partial H^{(\alpha)}_{\beta-1,l}}{ \partial A_j}+
\delta_{jl} F^{(\alpha-1)}_{\beta-1,l} + A_l \frac{ \partial H^{(\alpha-1)}_{\beta-1,l}}{ \partial A_j} \label{rec}\\
&& \frac{ \partial H^{(\alpha)}_{\beta,l}}{ \partial A_l}=\frac{ \partial H^{(\alpha-1)}_{\beta,l}}{ \partial B_l} \label{id}
\end{eqnarray}
and using the inductive hypothesis, one obtains by straightforward computation
\begin{displaymath}
P d H^{(\alpha-1)}_{\beta,l}-(R -(l-2) P) d H^{(\alpha)}_{\beta,l}=0.
\end{displaymath}
The case $\alpha=\beta$ can be easily verified by direct computation.

Notice that identities (\ref{iid}) and (\ref{iid2}) follow from~\rref{Fm} and \rref{Hm}. 
The identity (\ref{rec}) 
follows from the recursive formula
for $F^{(\alpha)}_{\beta,l}$:
\begin{eqnarray*}
&& F^{(\alpha)}_{\beta,l}=B_l F^{(\alpha)}_{\beta-1,l}+A_l F^{(\alpha-1)}_{\beta-1,l}\\
&& F^{(0)}_{0,l}=1\\
&& F^{(\alpha)}_{\beta,l}=0 \qquad {\rm if} \ \alpha>\beta \  {\rm or} \ \alpha<0,
\end{eqnarray*}
while (\ref{id}) can be proven again by induction. For $\alpha=2$ it holds:
\begin{displaymath}
\frac{ \partial H^{(2)}_{2,l}}{ \partial A_l}=2 A_l=\frac{ \partial H^{(1)}_{2,l}}{ \partial B_l} \qquad 
\frac{ \partial H^{(1)}_{2,l}}{ \partial A_l}=2 B_l=\frac{ \partial H^{(0)}_{2,l}}{ \partial B_l}
\end{displaymath}
Then, if $\alpha \leq \beta-1$ we can use the inductive hypothesis and get:
\begin{displaymath}
\frac{ \partial H^{(\alpha)}_{\beta,l}}{ \partial A_l}-\frac{ \partial H^{(\alpha-1)}_{\beta,l}}{ \partial B_l}= B_l \left( \frac{ \partial H^{(\alpha)}_{\beta-1,l}}{ \partial A_l}-\frac{ \partial H^{(\alpha-1)}_{\beta-1,l}}{ \partial B_l} \right)+ A_l \left(\frac{ \partial H^{(\alpha-1)}_{\beta-1,l}}{ \partial A_l}-\frac{ \partial H^{(\alpha-2)}_{\beta-1,l}}{ \partial B_l} \right)=0.
\end{displaymath} 
The case $\alpha=\beta$ is again a matter of simple computation.
\endpf
\begin{prop}\label{prop:5}
The Hamiltonians
\begin{equation*}\begin{split}
& K^{(\beta-k)}_{\beta,l}=\sum_{j=0}^{k-1}
{{k-1}\choose{j}}
(l-2)^{k-j-1}  H^{(\beta-j-1)}_{\beta,l} \\
& K^{(\beta)}_{\beta,l}=H^{(\beta)}_{\beta,l}\>,\qquad l=2,\dots,N\end{split}
\end{equation*}
satisfy the standard Lenard-Magri relations $
P d K^{(\alpha-1)}_{\beta,l}=R d K^{(\alpha)}_{\beta,l}, \quad \alpha=1,
\ldots\beta, \forall\> \beta.$
\end{prop}
{\bf Proof:} Using Lemma \ref{lemma:4}, we have:
\begin{eqnarray*}
&& \ \ \ R d K^{(\beta-k)}_{\beta,l}=\\
&&=P d \left( \sum_{j=0}^{k-1}{{k-1}\choose{j}}
\big((l-2)^{k-j-1}  H^{(\beta-j-2)}_{\be,l} + 
(l-2)^{k-j}  H^{(\beta-j-1)}_{\beta,l}\big) \right)=\\
&&= P d \left( K^{(\beta-k-1)}_{\beta,l}+\sum_{j=1}^{k-1}
\left[
{{k-1}\choose{j-1}}+{{k-1}\choose{j}}
\right]
(l-2)^{k-j}  H^{(\beta-j-1)}_{\beta,l} \right)=\\
&&= P d \left(  \sum_{j=0}^{k}
{{k}\choose{j}}
(l-2)^{k-j}  H^{(\beta-j-1)}_{\beta,l} \right) = 
P  d K^{(\beta-k-1)}_{\beta,l}  
\end{eqnarray*}
\endpf

\subsection{ Complete integrability for
$\boldsymbol{\alg}=\mathbf{sl(r)}$}\label{sect:c.i.}
We  now prove that the Hamiltonians (\ref{Hm}) 
together with  additional integrals one can recover from the 
global $SL(r)$ invariance of the model
provide complete integrability in the case $\alg=sl(r)$.  
We start by remarking that the content of Lemma~\ref{lemma:4} 
and Proposition~\ref{prop:5} can be rephrased as follow:
If we introduce the $N$ matrices:
\begin{equation}\label{eq:lmmus}
L_1=A_1,\qquad L_a=\la A_a+B_a, \quad a=2, \ldots,N
\end{equation}
then they evolve isospectrally along any of the vector field of the hierarchy, that is, 
(since the matrices $A_i$ are generically simple) along Lax type equations. 

The dimension of the
manifold $M= sl(r)^N$ is $d_M=(r^2-1)N$, and the dimension of the generic
symplectic leaf of $P$ is $d_S=d_M-N(r-1)=r(r-1)N$. We notice that the we
recover (as expected) all the Casimirs of $P$ considering: a) the spectral
invariants of $L_1=A_1$ (this gives $N-1$ {\em common} Casimirs), and b)
the higher order terms in expansions of $L_l$: $\Ha{\al}_{\al,l},\  \al=2,\ldots,r,\  l=2,\ldots, N$. 
Since it holds $H^{(0)}_{\beta,l}= \sum_{k=0}^{\beta} H^{(k)}_{\beta,l-1}$, we consider the set
\begin{displaymath}
H^{(\al)}_{\beta,l} \quad \beta=2,\dots,r \quad \alpha=1,\dots,\beta-1.
\end{displaymath}
This provides us
with a distinguished sequence 
of $\frac{r(r-1)}{2}$ mutually
commuting Hamiltonians. Clearly, if $l \neq l'$, the sets $ \{H^{(\al)}_{\beta,l}\}$ and
$\{H^{(\al)}_{\beta,l'}\}$ are functionally independent, since they depend on a 
different set of variables. 
So, the counting of the number of independent Hamiltonians
boils down to compute the counting of independent coefficients in the 
determinant
\[
\text{det}(\mu-\la A+B), A,B\in sl(r).
\]
This problem was solved in \cite{Du85}, (see, also, \cite{AHH}) and, actually, 
the number is exactly  
$\frac{r(r-1)}{2}$. Hence, the Lenard sequences associated with $R-\la P$
provide us with a total number of $(N-1) \frac{r(r-1)}{2}$ commuting
Hamiltonian, plus the $N(r-1)$ Casimirs.
For complete integrability we are missing $r(r-1)/2$ more commuting 
integrals. 

They are associated with the global $SL(r)$ invariance of the problem, and, in
the \bih\ picture, can be described as follows. For every $\tau\in sl(r)$ we
can consider the linear function 
\begin{displaymath}
H_\tau={\rm{tr}}(\sum_{i=1}^N A_i\tau).
\end{displaymath}
The Lenard ``sequence'' associated with such functions is somewhat peculiar;
indeed, since $R dH_\tau=(N-1) P dH_\tau$, we can associate to each $H_\tau$ a
Lenard diagram which is (up to a constant) a closed loop, 
to be compared with the usual ladder
typical of iterable Hamiltonians. 
However, the usual argument of \bih\ recurrence, shows that, for any $\tau$,
\[
\{K^{(\al)}_{\beta,l},H_\tau\}_P=\{K^{(\al)}_{\beta,l},H_\tau\}_R=0, \> \forall\> \al,\beta,l.
\]
Indeed, one has, e.g., the equality:
\[
\{K^{(\al)}_{\beta,l},H_\tau\}_P=\{K^{(\al+1)}_{\beta,l},H_\tau\}_R=(N-1)\cdot\{K^{(\al+1)}_{\beta,l},H_\tau\}_P.
\]
This argument shows how to recover, in the \bih\ formalism, the integrals
associated with the global $SL(r)$ invariance of the model. Clearly, this
family of $r^2-1$ integrals is not a commutative one.

To recover the maximal Abelian subalgebra inside the Poisson algebra
generated by the functions $H_\tau$, one can consider (see, e.g., \cite{DOF}):
\begin{description}
\item[a)] The $r-1$ independent 
elements $H_{h_1},\ldots,H_{h_{r-1}}$ associated with, say, the standard 
Cartan subalgebra of $sl(r)$;
\item[b)] The Gel'fand-Cetlin invariants, that is, 
the Casimirs of the nested subalgebras
\begin{equation}\label{eq:gece}
sl(2) \subset sl(3) \subset \dots \subset sl(r), 
\end{equation}
under the map 
$sl(r)^N\to sl(r)$ sending the $N$-tuple $\{A_1,\ldots,A_N\}$ into the total
sum, $A_{tot}=\sum_{i=1}^N A_i$.
\end{description}
Noticing that the Gel'fand-Cetlin functions corresponding to the last element
of the chain~\rref{eq:gece}
are given by $\sum_{k=0}^\beta H^{(k)}_{\beta,N}$,
we obtain
$r-1+ \sum_{i=2}^{r-1} (i-1)=\frac{r(r-1)}{2}$
additional commuting integrals, 
which is exactly the number of commuting integrals we were looking for to
insure complete integrability of the model.

We end this Section with a comment concerning super--integrability of the
model. To this end we remark that we have at our disposal two Poisson pencils
to construct families of commuting integrals for the Gaudin (homogeneous) 
Hamiltonian $H_G$: the pencil $R-\la P$ and the pencil $Q-\la P$, described in
Section~\ref{sec:general}\footnote{With the proviso in mind that one has to
  set $\sigma=0$ in those formulas}. 
On the $d_{N,r}=N(r(r-1))$--dimensional generic 
symplectic leaves of $P$ they give rise to two distinct $d_{N,r}/2)$
families of integrals of the motion $K_m^{lj}$ and $\widetilde{K}^{lj}_m$.
Direct computations (which we performed for $r=3,4$ and $N\le 6$) suggest 
that the number of functionally independent elements in the union of the two
families be  $d_{N,r}-(r-1)$. 
In other words, also taking into account the integrals coming from the global
$SL(r)$ invariance of the model, we have super-integrability for the $sl(r)$
Gaudin model, that, however, is maximal only for the $sl(2)$ case.

\section{Separation of Variables for the $\mathbf{sl(2)}$ case}\label{sect:sepslr} 
We consider now the $N$-particle $sl(2)$ case. The aim is to show that the
Hamilton--Jacobi equations associated with the Hamiltonians
\begin{equation}
  \label{eq:s1}
  H_a=\text{Tr}\big(A_i\cdot\sum_{j=1}^{a-1}A_j\big),\quad a=2, \ldots, N,
\end{equation}
and, in particular, the H--J equations associated with the physical 
Hamiltonian $H_G=\frac12 \sum_{i=1}^N H_i$ are separable in a very ``simple''
set of coordinates. 
Our analysis is based on the so--called  \bih\ scheme for SoV, 
recently introduced in the literature (see, e.g., \cite{MT,Bl,FMP}). 
In particular, we will use the results for systems with an 
arbitrary number of (anchored) Lenard chain exposed in \cite{FP}.

We consider the manifold $M=sl(2)^N$, endowed with 
the Poisson pencil $R-\la P$,
explicitly parametrized with the $N$ matrices
\begin{equation}
A_i=\left[\begin{array}{cc} h_i& f_i\\ e_i&-h_i\end{array}\right].
\end{equation}
The generic symplectic leaf $S$ of $P$ is a  
$2N$ dimensional symplectic manifold, defined by the equations
\begin{displaymath}
C_i=\frac12\text{Tr}A_i^2=h_i^2+e_i f_i, \quad  i=1,\ldots, N,
\end{displaymath}
and can be (generically) endowed with the $2N$ coordinates $\{h_i,
f_i\}_{i=1,\ldots, N}$.

According to the \bih\ scheme~\cite{FP0}, one modifies
the tensor $R$ in order to obtain a second compatible structure on $S$.
Let us define the $N$ vector fields
\begin{equation}\label{eq:s3}
Z_i=\frac{1}{f_i}\ddd{}{e_i},
\end{equation}
and, for any pair of functions $F,G$ on $M$,  
the following brackets:
\begin{equation}
  \label{eq:s2}
  \{F,G\}_{R'}=\{F,G\}_{R}-\sum_{a=2}^N \big(\{F,H_a\}_P
Z_a(G)-\{G,H_a\}_P
Z_a(F)\big),
\end{equation}
where  $X(F)$ denotes the Lie derivative of $F$ with 
respect to the vector field $X$. 

Thanks to the specific form of the 
vector fields $Z_a$, these new brackets restrict to $S$ and give 
rise to {\em Poisson brackets} on
$S$, which are compatible with the ones naturally induced by 
$P$.
So,
$S$ is  a symplectic manifold with respect to the
restriction of the brackets associated with $P$, and is endowed with a
$(1,1)$ tensor $\CN$,  with vanishing \Nij\ torsion defined by
\[
\CN=R'\circ P^{-1}.
\]
In such a geometrical setting, the \bih\ scheme for SoV considers sets of
coordinates $\{u_i,v_i\}$ 
(called \Nij\ coordinates) associated with the eigenvalues
$\la_i$ of $\CN$, characterized by the equations:
\begin{equation}
    \label{eq:b3}
    \CN^* du_i=\la_i du_i,\quad \CN^* dv_i=\la_i dv_i,
  \end{equation}  
whose Poisson brackets attain the remarkable form~\cite{Magri,GZ}:
\begin{equation}\label{eq:s4}
\begin{split}
&\{u_i,u_j\}_P=\{u_i,u_j\}_{R'}=\{v_i,v_j\}_P=\{v_i,v_j\}_{R'}=0\\
&\{u_i,v_j\}_P=\delta_{ij} \vartheta_i(u_i,v_i),\quad  
\{u_i,v_j\}_{R'}=\la_i\{u_i,v_j\}_P,
\end{split}
\end{equation} 
for some functions $\vartheta_i(u_i,v_i)$.

We shall prove the above statements directly displaying a set of 
\Nij\ coordinates on the symplectic leaves of $P$. 
\begin{prop}\label{prop:lambda} 
The $2N$ functions
\begin{equation}\label{eq:s5}\begin{split}
\la_1&=\sum_{i=1}^N f_i,\qquad 
\lambda_a= -\frac{ \sum_{k=1}^{a-1} f_k}{f_a}+(a-2), \qquad a=2,\dots, N\\
\mu_1&=\sum_{i=1}^N h_i,\qquad
\mu_a =\left(\lambda_a- (a-2)\right) h_a+ \sum_{k=1}^{a-1} h_k, \qquad a=2,
\dots, N
\end{split}
\end{equation}
provide a set of  \Nij\ coordinates on $S$. In particular the $\la_a,
a=2,\ldots,N$ are the non-vanishing eigenvalues of $\CN^*$, while $\la_1$ and
$\mu_1$ span its (2--dimensional) kernel.    
\end{prop}
{\bf Proof.}
According to Equation~\rref{eq:s2} and the definition 
of the \Nij\ tensor $\CN$, 
noticing that both $P$ and $R'$ restrict to $S$ and that
$Z_a(\la_i)=Z_a(\mu_i)=0$ for $a=2, \ldots, N, i=1,\ldots, N$,
we have to show that,  for any coordinate $x_i$,  
(that is, $x_i=e_i,h_i,f_i,\> i=1,\ldots, N$)
it holds
\begin{equation}
\begin{split}
&\{ x_i, \lambda_1 \}_R- \sum_{a=2}^N \{ H_a, \lambda_1 \}_P Z_a(x_i)=0\\
& \{ x_i, \lambda_b \}_R- \sum_{a=2}^N \{ H_a, \lambda_b \}_P Z_a(x_i) -
 \lambda_b \{ x_i, \lambda_b \}_P ,\> b=2,\ldots,N
\end{split} \label{lambda}
\end{equation}
as well as
\begin{equation}
\begin{split}
&\{ x_i, \mu_1 \}_R- \sum_{a=2}^N \{ H_a, \mu_1 \}_P Z_a(x_i)=0\\
& \{ x_i, \mu_b \}_R- \sum_{a=2}^N \{ H_a, \mu_b \}_P Z_a(x_i) =
 \lambda_b \{ x_i, \mu_b \}_P ,\> b=2,\ldots,N.
\end{split} \label{mu}
\end{equation}
The proof that these equations holds true is a matter of direct
computations. One simply has to plug
the explicit expressions of the
Poisson brackets
\begin{eqnarray*}
&& \{ h_i, e_j \}_P= \delta_{ij} e_j,\quad
 \{ h_i, f_j \}_P= -\delta_{ij} f_j,\quad
 \{ e_i, f_j \}_P= 2 \delta_{ij} h_j\\
&& \{ h_i, e_j \}_R= \delta_{ij} \left[ (i-1) e_i- \sum_{k=1}^{i-1} e_k
\right] + 
\theta_{(i-j)}e_j +  \theta_{(j-i)} e_i\\
&& \{ h_i, e_j \}_R= -\delta_{ij} \left[ (i-1) f_i- \sum_{k=1}^{i-1} f_k
\right] - 
\theta_{(i-j)}f_j -  \theta_{(j-i)} f_i\\
&& \{ e_i, f_j \}_R= 2 \left\{ \delta_{ij} \left[ (i-1) h_i- \sum_{k=1}^{i-1}
    h_k \right] + 
\theta_{(i-j)}h_j +  \theta_{(j-i)} h_i \right\}.
\end{eqnarray*}
into equations~\rref{lambda} and \rref{mu}, and
use the identities
\begin{eqnarray}
&& \sum_{j=1}^{n-1} \delta_{ij}= \theta_{(n-i)}, \quad 
\sum_{j=1}^{n-1} \theta_{(j-i)} 
F_j= \theta_{(n-i)} \sum_{j=i+1}^{n-1} F_j \label{uno}\\  
&& \sum_{j=1}^{n-1} \theta_{(i-j)}F_j=(\theta_{(i-n)}+\delta_{in})
\sum_{j=1}^{n-1} F_j+ \theta_{(n-i)} \sum_{j=1}^{i-1} F_j \label{due}.
\end{eqnarray}
For example, let us consider 
$x_i\equiv h_i$. Since  $Z_a({h_i})=0$, we have (for $n\geq 2$):
\begin{displaymath}
\{ h_i, \lambda_n \}_P= (\la_n-n+2) 
\delta_{in}+ \theta_{(n-i)} \frac{f_i}{f_n},
\end{displaymath} 
and
\begin{eqnarray*}
&& \{ h_i, \lambda_n \}_R= \frac{1}{f_n} \sum_{j=1}^{n-1} \left\{ \delta_{ij} \left[ (i-1) f_i- \sum_{k=1}^{i-1} f_k \right] + 
\theta_{(i-j)} f_j +  \theta_{(j-i)} f_i \right \}+\\
&& + \frac{\la_n-n+2}{f_n} \left\{ \delta_{in} \left[ (i-1) f_i- \sum_{k=1}^{i-1} f_k \right] + 
\theta_{(i-n)} f_n +  \theta_{(n-i)} f_i \right \}=\\
&& =\lambda_n \left( \delta_{in}  (\la_n-n+2)+ \theta_{(n-i)} \frac{f_i}{f_n} \right)
\end{eqnarray*}
thanks to identities (\ref{uno}), (\ref{due}).

The other cases of equations~(\ref{lambda},\ref{mu}) are proved 
with similar computations.
\endpf
To construct a set of {\em canonical} \Nij\ coordinates (usually considered in
the \bih\ scheme for SoV and quite naturally termed {\em \dncoo})
$\{\la_i,\phi_i\}_{i=1,\ldots, N}$ from the \Nij\ coordinates
$\{\la_i,\mu_i\}_{i=1,\ldots,N}$ one notices that 
an explicit computation gives:
\begin{equation}
  \label{eq:s19}
\{\la_1,\mu_1\}=-\la_1, \qquad \{\la_a,\mu_a\}=(\la_a-(a-2))(\la_a-(a-1)),
a=2,\ldots, N.    
\end{equation}
Indeed, the first equation is trivially verified; for the remaining set of
$N-1$ relations one has
\begin{eqnarray*}
&& \{ \mu_n, \lambda_n \}= - \frac{1}{f_n}  \{ \sum_{i=1}^{n-1} h_i,  
\sum_{i=1}^{n-1} f_i \} - (\la_n-(n-2)) \{ h_n, 1/f_n \} \sum_{i=1}^{n-1} f_i=\\
&& =\frac{\sum_{i=1}^{n-1} f_i}{f_n}- (\la_n-(n-2)) \frac{\sum_{i=1}^{n-1} f_i}{f_n}= 
(\la_n-(n-2))(\la_n-(n-1)).
\end{eqnarray*}
Hence, one can choose 
\begin{equation}\label{eq:canco}
\phi_1=-\frac{\sum_{i=1}^{N} h_i}{\sum_{i=1}^{N} f_i},\qquad
 \phi_a= \frac{\mu_a}{(\la_a-(a-2))(\la_a-(a-1))} \quad a=2, \dots,N
\end{equation}
to have, together with the $\la_i$ a set of \dncoo.

To find the separation relations, we make contact with the so--called 
Sklyanin magic recipe~\cite{Sk95}.  To this end, we modify the Lax 
matrices~\rref{eq:lmmus}, by a suitable shift in the spectral parameter $\la$.
Namely we define
\begin{equation}\label{eq:lmgf}
\wid{L}_1=L_1;\quad \wid{L}_a=(\la-(a-2))A_a+\sum_{b=1}^{a-1} A_b, \quad a=2, \ldots, N.
\end{equation}
We notice that the spectral invariants 
of $\wid{L_a}$ are combinations of the 
Hamiltonians $\Ha{\al}_{\be, l}$ and of the common Casimirs
we considered in Section \ref{sect:c.i.}, and provide an 
equivalent set of involutive constants of the motion (together with $H_1=\sum h_i$).

As one can easily notice, the
\Nij\ coordinates $\la_a, a=2,\ldots, N$ are nothing but the zeroes of the
$(1,2)$ entry of the Lax matrix $\wid{L}_a$ of eq.~\rref{eq:lmgf}, while the
\Nij\ coordinates $\mu_a, a=2,\ldots, N$ are the values for $\la=\la_a$ of
the $(1,1)$ entry of the same matrix.
Taking into account that $\mu_1$
is the first Hamiltonian $H_1$, we see that the \Nij\ coordinates, the
Hamiltonians and the Casimirs $C_i=\text{Tr}(A_i)^2$ are related 
by the separated equations:
\begin{equation}\label{eq:pane}
\mu_1-H_1=0;\qquad
\text{Det}\Big(\mu_a-\big((\la_a-(a-2))A_a+\sum_{i=1}^{a-1}A_i\big)\Big)=0,
\end{equation}
whence one can directly find, using the relations ~\rref{eq:canco}, 
the ``canonical'' separation relations for the Hamilton-Jacobi equation associated
with the $sl(2)$ Gaudin Hamiltonians. We notice that these are quadratic equations 
in the separated coordinates, and hence explicitly solvable by elementary functions,
for every number of sites $N$.
\section{Summary of the Results}
In this paper we have used tools from \bih\ geometry to study the
integrability of  (rational XXX) Gaudin models, associated with the Lie
algebra $sl(r)$. We first framed the general (that is, inhomogeneous) 
model within the
Gel'fand--Zakharevich scheme by selecting a suitable pencil of Poisson
brackets induced by a natural family of Poisson brackets on the space of
matrix polynomials. 

Then we extensively studied the {\em homogeneous case}. 
We considered an alternative complete set of mutually commuting 
constants of the motion $I_k$ 
(independent of the parameters  usually entering the formulation of
Gaudin models). These integrals (actually, a subfamily thereof), 
in the $su(2)$ case, coincide with
the Hamiltonians of the bending flows of Kapovich and Millson on  
the moduli space of polygons in the Euclidean space.

We introduced an ``additional'' Poisson tensor which forms, together with the
standard Lie--Poisson tensor, a \bih\ pencil.
The GZ analysis of such a \bih\ structure
provides exactly the alternative set of constants of the motion $I_k$ 
for the $sl(2)$ case.
By using such a  \bih\ scheme, 
we extended this analysis to the $sl(r)$ case; in 
particular, we show that the higher rank counterparts of the additional set of 
integrals guarantee complete integrability of the $sl(r)$--Gaudin
magnet. Since, we still have at our disposal the Jur\v{c}o--Sklyanin
integrals, we conclude that the Gaudin magnet is super--integrable (although 
we could not establish {\em maximal super--integrability}) also in the  $sl(r)$
case. 

We furthermore have explicitly shown in the $sl(2)$ case that the
Hamilton--Jacobi equations associated with the set of additional integrals can
be solved by separation of variables, using the \bih\ scheme for SoV which has 
recently been considered in the literature. Actually, what we found is a 
set of separation coordinates {\em alternative} to the ``standard'' one found
by Sklyanin and the ``Montreal group'', based on the standard Lax
representation for the (homogeneous) Gaudin model. This should not be
regarded as a surprise, since it is well known that superintegrability is
related with the existence of different sets of separation
coordinates. In this set of coordinates, the H-J equations 
can be explicitly solved by elementary functions, or, 
otherwise stated, the separation coordinates (for the $sl(2)$-magnet) 
live on genus $0$ spectral curves for {\em any} number of particles, 
while, in the  ``standard picture'', the genus of the spectral 
curve grows linearly with $N$.

\section*{Acknowledgments} We wish to thank J. Harnad for 
many useful discussions, and, especially, for drawing our attention to 
the papers~\cite{KM,FlMi01} concerning the bending flows. Also, 
we thank B. Dubrovin, F. Magri, M. Pedroni, and O. Ragnisco for their interest
in this work.

\newappendix{}

\noindent In this appendix we sketch the proof of Lemma 1.
Actually we will  prove the converse statement, i.e., that under the 
map \rref{map} the Poisson tensor \rref{PDB} is sent exactly in the diagonal
Poisson tensor \rref{PDA}.

We denote with $J$ the Jacobian of the transformation:
\begin{equation}
J_{ij}= \frac{\partial B_{i-1}}{\partial A_j}=(-1)^{N-i}
s_{N-i}(a_1,\dots,\hat{a_j},\dots,a_N). \label{Jacobian}
\end{equation}

Using the identity:
\begin{equation}
\sum_{j=1}^N x^{j-1} (-1)^{N-j} s_{N-j}(a_1,\dots,\hat{a_k}, \dots,a_N)=\prod_{l \neq
k} (x-a_k) \label{useful}
\end{equation}
the inverse matrix of \ref{Jacobian} is easily obtained:
\begin{equation}
(J^{-1})_{ij}=\frac{a_i^{j-1}}{\prod_{k \neq i} (a_i-a_k)}.
\end{equation}
We have:
\begin{equation}\label{A6} (J^{-1} P J^{-1})^t)_{in}=
 \frac{(-1)^N}{\prod_{m \neq i}(a_i-a_m) \prod_{p \neq n}(a_n-a_p)}
(P_{in}^{(1)}-P_{in}^{(2)})
\end{equation}
with: 
\begin{eqnarray}
&& P_{in}^{(1)}=\sum_{r=0}^{N-1}  \sum_{l=r+1}^N (-1)^l s_{N-l}(a_1,\dots,a_N)
\sum_{k=r+1}^l a_i^{r+l-k} a_n^{k-1}[B_r, \cdot] \label{primo}\\
&& P_{in}^{(2)}=\sum_{r=1}^{N}  \sum_{l=0}^{r-1} (-1)^l s_{N-l}(a_1,\dots,a_N)
\sum_{k=l+1}^r a_i^{r+l-k} a_n^{k-1}[B_r, \cdot]. \label{secondo}
\end{eqnarray}
Subtracting \rref{primo} and \rref{secondo},
by using induction and the identities
\begin{eqnarray}
&& s_i(a_1, \dots,a_{N+1})=s_i(a_1,\dots,a_N)+a_{N+1}s_{i-1}(a_1,\dots,a_N)
\label{id2} \\ 
&& s_i(a_1,\dots,a_N)=0 \ {\rm{if}} \ i<0 \  {\rm{or}} \ i>N \nonumber\\
&& \sum_{l=0}^{N} (-1)^l s_{N-l}(a_1,\dots,a_{N})
x^{l}=(-1)^N \prod_{i=1}^N(x-a_i)  \label{fundamental}
\end{eqnarray}
one proves that the coefficient of $B_r$ in formula (\ref{A6}) vanishes if
$i \neq n$. 

Now let us consider the diagonal terms.
\begin{displaymath}
P^{(1)}_{ii}-P^{(2)}_{ii}=\sum_{r=0}^N \left(a_i^r \sum_{l=0}^N (-1)^l s_{N-l}(a_1,\dots,a_N)(l-r) a_i^{l-1}[B_r, \cdot]\right).
\end{displaymath}
Since 
\begin{displaymath}
 \sum_{l=0}^N (-1)^l s_{N-l}(a_1,\dots,a_N) \, l \, a_i^{l-1}= 
\left. (-1)^N \frac{d}{d x} \left( \prod_{j} (x-a_j) \right) \right|_{x=a_i}=
(-1)^N \prod_{j \neq i} (a_i-a_j)
\end{displaymath} 
we get (using (\ref{map}) and (\ref{useful})) :
\begin{displaymath}
P^{(1)}_{ii}-P^{(2)}_{ii}=(-1)^N \prod_{j \neq i} (a_i-a_j) \sum_{r=0}^N 
a_i^r [B_r, \cdot]
= (-1)^N \left(\prod_{j \neq i} (a_i-a_j)\right)^2 [A_i, \cdot]
\end{displaymath}
whence the assertion.
\endpf

\newappendix{}

\begin{lemma}
The Poisson tensors $Q$ \rref{P1} and $R$ \rref{P2} are not compatible 
for $N \geq 3$ for any choice of the
parameters $a_1,\dots,a_N$.
\end{lemma}
{\bf Proof:} An explicit computation shows that
the Poisson tensor $Q$ can be written 
in the ``coordinates'' $\{A_1, \ldots, A_N\}$  as:
\begin{displaymath}
\{ F, G \}_Q=\sum_{i,j,k} q_{ijk} \text{Tr}
\left(A_k \left[ \frac{\partial G}{\partial A_j},
\frac{\partial F}{\partial A_i}\right] \right),
\end{displaymath}
with:
\begin{eqnarray*}
&& q_{ijk}=(-1)^N \left\{ \delta_{ij} \left( \xi_j \delta_{jk}+ \beta_j \frac{(1-\delta_{jk})}{\eta_{jk}} \right)+ \frac{(\beta_i \delta_{jk}- \beta_j
\delta_{ik})}{\eta_{ji}} \right\}\\
&& \eta_{ij}=\left\{
\begin{array}{ccc}
a_i-a_j & {\rm if} & i \neq j\\
1  & {\rm if} & i=j 
\end{array} \right.
\qquad \beta_i= \prod_{k \neq i} \frac{a_k}{\eta_{ik}} \qquad \xi_i=\sum_{j\neq i} \frac{\beta_j}{\eta_{ji}}.
\end{eqnarray*}
Using this expression is easy to evaluate  the Schouten bracket of $Q, \ R$ on the differentials
of the functions
\begin{displaymath}
F=\text{Tr}(A_1 h) \qquad G=\text{Tr}(A_2 x) \qquad H=\text{Tr}(A_2 h)
\end{displaymath}
where with $x$ and $h$ we denoted two constant matrices satisfying $[h,x]=x$.
We have:
\begin{eqnarray}
&& \back \back \back  [Q, R]_S(d F, d G, d H)= \nonumber\\
&& \back \back \back =(-1)^N \left[ \xi_2- \xi_1 + \frac{(\beta_1-\beta_2)}{\eta_{21}}+ 
\beta_2 \sum_{j=3}^N \frac{1}{\eta_{2j}} \right] tr(A_1 x)+
\beta_1  \sum_{k=3}^N \frac{1}{\eta_{k1}} tr(A_k x) \label{incomp}
\end{eqnarray}   
A necessary condition for (\ref{incomp}) to vanish 
is that $\beta_1=0$, i.e., one of the constants
$a_2,\dots,a_N$ must  be equal to zero.
Let us suppose $a_k=0, \> k>2$; then
\begin{displaymath}
[Q, R]_S(d F, d G, d H)= (-1)^N\left( \frac{1}{a_2} -  \frac{1}{a_1} \right) \text{Tr}(A_1 x) \neq 0
\end{displaymath}
since $a_1\neq a_2$.

In the case $a_2=0$, we have instead:
\begin{equation}
[Q, R]_S(d F, d G, d H)=(-1)^N \text{Tr}(A_1 x) \sum_{j=3}^N \frac{1}{a_j}. \label{Neq3}
\end{equation}
If $N=3$ then (\ref{Neq3}) is different from zero, so we can assume $N>3$. 
But if $a_2=0$ and 
$N>3$ we can consider
\begin{displaymath}
F'=\text{Tr}(A_1 h) \qquad G'=\text{Tr}(A_2 x) \qquad H'=\text{Tr}(A_3 h).
\end{displaymath}
Since
\begin{displaymath}
[Q, R]_S(d F', d G', d H')=(-1)^{N+1} \frac{1}{a_3} tr(A_1 x) \neq 0
\end{displaymath}
the proof is concluded.


\begin{thebibliography}{99}

\bibitem{AHH} M. Adams, J, Harnad, J. Hurtubise, 
\textsl{Darboux Coordinates and Liouville-Arnold Integration 
in Loop Algebras}, \cmp{155}{385--413} {(1993)}\\
 M. Adams, J, Harnad, J. Hurtubise, 
\textsl{Darboux Coordinates on Coadjoint orbits of Lie Algebras},
\lmp{40}{41--57}{(1996)}.

\bibitem{Bl} M. B\l aszak, \textsl{On separability of bi--Hamiltonian chain
with degenerated Poisson structures}. \jmp{39}{3213--3235}{1998}

\bibitem{Balrag} A. Ballesteros, 
O. Ragnisco, \textsl{Classical Hamiltonian systems with
$sl(2)$ coalgebra symmetry and their integrable deformations},  
\jmp{43}{954--969}{2002} 

\bibitem{Ragn} A. Ballesteros, M. Corsetti, O. Ragnisco,
\textsl{N-dimensional classical integrable systems from Hopf Algebras},
Czech. J.Phys. {\bf{46}}, 1153--1165 (1996).

\bibitem{DOF} Yu.A. Drozd, S.A. Ovsienko, V.M. Futorny, 
\textsl{On Gel' fand-Zetlin modules},
Proceedings of the Winter School on Geometry and Physics (Srni, 1990),  
Rend. Circ. Mat. Palermo (2) Suppl. No. {\bf{26}}, 143--147, (1991).

\bibitem{Du85} 
Dubrovin B.A. \textsl{Matrix finite-zone operators}, J. Sov. Math. {\bf{28}},
20--50 (1985).

\bibitem{FMP} G. Falqui, F. Magri, M. Pedroni, \textsl{Bihamiltonian geometry and separation of 
variables for Toda lattices}, J Nonlinear Math. Phys. {\bf{8}} Suppl., 118--127 (2001).

\bibitem{FP0} G. Falqui, M. Pedroni, \textsl{On a Poisson Reduction for 
Gel'fand-Zakharevich Manifolds}, (nlin.SI/0204050), Rep. of Math. Phys., {\bf 50}, 395--407 (2002).

\bibitem{FP} G. Falqui, M. Pedroni, \textsl{Separation of variables for 
bi-Hamiltonian systems}, nlin-SI/0204050, to appear in Math. Phys. Anal. Geom
(2003). 

\bibitem{FM1} G. Falqui, F. Musso, \textsl{Bi--Hamiltonian Geometries of Gaudin Models:
a Case Study}, Proceedings of the International Conference SPT 2002 Symmetry and 
Perturbation Theory, World Scientific, Singapore, 42--50 (2002). 

\bibitem{FlMi01} H. Flaschka, J. Millson, \textsl{The moduli space of weighted
configurations on projective space}, math.SG/0108191.

\bibitem{Gaudin1} M. Gaudin, \textsl{Diagonalisation d' une classe
d' hamiltoniens de spin}, J. de Physique {\bf{37}}, 1087--1098 (1976).\\ See, also,
M. Gaudin, \textsl{La Fonction d' Onde de Bethe}, Masson,
Paris (1983).

\bibitem{Gekhtman} M.I. Gekhtman, \textsl{Separation of variables in the classical ${\rm SL}(N)$ magnetic chain},
\cmp{167}{3}{593--605, (1995)}.

\bibitem{GZ} I. M. Gel'fand, I. Zakharevich,
{\em On the local geometry of a bi-Hamiltonian structure.}
In: The Gel'fand Mathematical Seminars 1990-1992
(L. Corwin et al., eds.), Birkh\"auser, Boston, 51--112 (1993).
\\   I. M. Gel'fand, I. Zakharevich,
{\em Webs, Lenard schemes, and the local geometry of \bih\ Toda 
and Lax structures}
Selecta Mathematica, new series, {\bf 6} 131--183 (2000).

\bibitem{HaYe03} J. Harnad, O. Yermolaeva, \textsl{Superintegrability, Lax matrices and separation of variables},
nlin.SI/0303009. 


\bibitem{Jurco} B. Jur\v{c}o, \textsl{Classical Yang-Baxter equations and 
quantum integrable systems}, \jmp{30}{1289--1293}{1989}

\bibitem{KM} M. Kapovich, J. Millson,  \textsl{The symplectic geometry 
of polygons in Euclidean space},J. Differ. Geom. {\bf{44}}, 479--513 (1996)

\bibitem{Kar} V. Karimipour, 
\textsl{Integrable Structure of the New Calogero Models}, 
\jmp{39}{913-920}{1998}.
 
\bibitem{MaMo84} F. Magri, C. Morosi, \textsl{Quaderno S 19/1984 del Dip. di Matematica
dell'Universit\`a di Milano}, (1984).
\bibitem{Magri} F. Magri, \textsl{Geometry and Soliton Equations}, Atti. Acc. Sci. Torino,
{\bf 124} (Suppl.), 181--209 (1990).

\bibitem{MT} C. Morosi, G Tondo, 
\textsl{ Quasi--Bi--Hamiltonian systems and separability.}
J. Phys. A: Math. Gen. {\bf 30} (1997), 2799--2806.

\bibitem{PV} M. Pedroni, P. Vanhaecke, \textsl{A Lie algebraic generalization of the Mumford system, 
its symmetries and its multi-Hamiltonian structure},  Regul. Chaotic Dyn. {\bf{3}}, 132--160, (1998).

\bibitem{RSTS} A.G. Reyman, M.A. Semenov-Tian-Shansky,
\textsl{Group-Theoretical Methods in the Theory of Finite-Dimensional
Integrable Systems} in \textsl{Dynamical Systems VII}, Springer (1994).
 
\bibitem{Scott} D.R.D. Scott, \textsl{Classical functional Bethe ansatz for ${\rm SL}(N)$: 
separation of variables for the magnetic chain}, \jmp{35}{5831--5843}{1994}

\bibitem{Sk92} E.K. Sklyanin \textsl{Separation of Variables in the Classical 
Integrable SL(3) Magnetic Chain}, Comm. Math. Phys. {\bf{150}}, 181--192 (1992) 

\bibitem{Sk89} E. K. Sklyanin, \textsl{Separation of Variables in the 
Gaudin Model}, J. Sov. Math. {\bf{47}}:2, 2473--2488 (1989).

\bibitem{Sk95}  E. K. Sklyanin, \textsl{Separation of Variables: new trends}
Progr. Theor. Phys. Suppl. {\bf 118} 35--60 (1995).

\end{thebibliography}
\end{document}